\documentclass[preprint,12pt,onecolumn,superscriptaddress,floatfix,showpacs]{revtex4-1}
\usepackage{graphicx}%
\usepackage{dcolumn}%
\usepackage{bm}%

\usepackage[utf8]{inputenc}  
\usepackage[T1]{fontenc}     
\usepackage[british]{babel}  
\usepackage[sc,osf]{mathpazo}
\usepackage[scaled=0.86]{berasans}  
\usepackage[colorlinks=true, citecolor=blue, urlcolor=blue]{hyperref}  
\usepackage{graphicx} 
\usepackage[babel]{microtype}  
\usepackage{amsmath,amssymb,amsthm,bm,amsfonts,mathrsfs,bbm} 

\usepackage{xspace}  
\usepackage{pgfplots}

\newcommand\id{\leavevmode\hbox{\small1\kern-3.3pt\normalsize1}}

\newcommand{\p}{{\bf p}}

\begin{document}
\title{Biased Non-Causal Game}

\author{Some Sankar Bhattacharya}
\email{somesankar@gmail.com}
\affiliation{Physics and Applied Mathematics Unit, Indian Statistical Institute, 203 B. T. Road, Kolkata 700108, India.}

\author{Manik Banik}
\email{manik11ju@gmail.com}
\affiliation{Optics \& Quantum Information Group, The Institute of Mathematical Sciences, C.I.T Campus, Tharamani, Chennai 600 113, India.}

\begin{abstract}
The standard formulation of quantum theory assumes that events are ordered is a background global causal structure. Recently in Ref.[\href{http://www.nature.com/ncomms/journal/v3/n10/full/ncomms2076.html}{Nat. Commun. {\bf3}, 1092 (2012)}], the authors have developed a new formalism, namely, the \emph{process matrix} formalism, which is locally in agreement with quantum physics but assume no global causal order. They have further shown that there exist \emph{non-causal} correlations originating from \emph{inseparable} process matrices that violate a \emph{causal inequality} (CI) derived under the assumption that events are ordered with respect to some global causal relation. This CI can be understood as a guessing game, where two separate parties, say Alice and Bob, generate random bits (say input bit) in their respective local laboratories. Bob generates another random bit (say decision bit) which determines their goal: whether Alice has to guess Bob's bit or vice-verse. Here we study this causal game but with biased bits and derive a biased causal inequality (BCI). We then study the possibility of violation of this BCI by inseperable process matrices. Interestingly, we show that there exist \emph{inseparable} qubit process matrices that can be used to violate the BCI for arbitrary bias in the decision bit. In such scenario we also derive the maximal violation of the BCI under local operations involving traceless binary observables. However, for biased input bits we find that there is a threshold bias beyond which no valid qubit process matrix can be used to violate the causal inequality under \emph{measurement-repreparation} type operation.            
\end{abstract}

\maketitle
\section{Introduction}\label{sec1}
The usual perception of the physical world relies on the notion that events are ordered with respect to some global time parameter. The standard formulation of quantum theory, likewise Newtonian physics, assumes such a global order. Interestingly, in recent time it has been shown that a larger set of correlations can be obtained under the assumption that local operations are described by quantum theory, but with no reference to any global causal relation between these operations \cite{Oreshkov'2012}. Such correlations compatible with indefinite causal order are called \emph{causally inseparable}. Oreshkov \emph{et al.} have developed an interesting new mathematical technique, namely, \emph{the process matrix formalism} to capture all such correlations that are locally in agreement with quantum physics \cite{Oreshkov'2012}. At deeper foundational level this study initiates one possible approach to formulate the theory of quantum gravity which should be a \emph{probabilistic} theory equipped with a \emph{dynamic} space-time \cite{Hardy'2005,Hardy'2007}. During recent time it has generated a lot of research interests \cite{Chiribella'2012,Baumeler'2013,
Araujo'2014,Brukner'2014,Ibnouhsein'2014,Oreshkov'2014,
Cerf'2015,Oreshkov'2015,Araujo'2015,Branciard'2015,
Feix'2015,Brukner'2015}. 

Operationally the non-causal correlations are well explained in terms of causal games\cite{Branciard'2015}. One such game has been introduced in the seminal paper by Oreshkov et. al.\cite{Oreshkov'2012}. For our purpose we consider this game where two parties, (say) Alice and Bob, reside in separate laboratories that are completely shielded from the rest of the world. At each run of the game, system enters and exits each laboratory only once. Each of Alice and Bob, respectively, generates random bits denoted by $a$ and $b$, with $a,b\in\{0,1\}$, let call these input bits or in short I-bits. Bob generates another random bit $b'\in\{0,1\}$ (decision bit or D-bit), which determines their action: if $b'=0$, Alice has to guess Bob's bit $b$, whereas if $b'= 1$, Bob has to guess $a$. Their goal is to maximize the probability of success, $p_{succ}:=\frac{1}{2}[p(x=b|b'=0)+p(y=a|b'=1)]$, where $x$ and $y$ denote Alice's and Bob's guess, respectively. It has been shown that if events are ordered with respect to some global causal relation then under any strategy followed by Alice and Bob the success probability satisfies the following inequality \cite{Oreshkov'2012},
\begin{equation}\label{causal}
p^{csl}_{succ}\le\frac{3}{4}.
\end{equation} 
However, if the local laboratories are described by quantum mechanics, but no assumption about a global causal structure is made, it is in principle possible to violate the causal inequality (\ref{causal}) giving rise to the concepts of non-causal correlations. To capture such correlations the authors introduce the process matrix formalism which generalizes the standard quantum formulation. 

In this work we study the above causal game, but with bias in choosing the I-bits as well as the D-bit. Unlike the original game, we assume that the I-bits and D-bit are not uniform, rather have bias $\alpha$ and $\beta$, respectively, with $\frac{1}{2}\le\beta,\alpha<1$. Such a biased scenario is specified by the pair $(\alpha,\beta)$, with $(\frac{1}{2},\frac{1}{2})$ denoting the original unbiased case. For arbitrary $(\alpha, \beta)$ we derived biased causal inequality (BCI) which is satisfied whenever events obey some deterministic as well as probabilistic global causal order. Then we show that for the case $(\frac{1}{2},\beta)$, i.e., when I-bits are completely random but D-bit is biased, there always exist (inseparable) qubit process matrices giving rise to non-causal correlations that violate the causal inequality. In this case we also find the maximal violation of the causal inequality under a restricted set of local operations involving traceless binary observables. However, for the case  $(\alpha,\frac{1}{2})$ with $\alpha$ exceeding a threshold value we find that no qubit process matrix can be used to violate the causal inequality under \emph{measurement-repreparation} type operation. 

Organization of the paper is as follows: in section-(\ref{sec2}) we discuss the biased causal game and derive the BCI; in section-(\ref{sec3}) we present a brief description of process matrix formalism and in section-(\ref{sec4}) give example of causal inseparable process matrices that violate the BCI in particular case. Section-(\ref{sec5}) contains the maximal violation of this BCI under a restricted set of local operations and then we present our discussions in section-(\ref{sec6}). 

\section{Biased causal inequality}\label{sec2} 
In the introduction we have already discussed the setting of the causal game. The causal inequality (\ref{causal}) has been derived under the following assumptions:
\begin{itemize}
\item[(i)] \emph{Definite causal structure}: all the relevant events occuring in the game are localized in a causal structure; 
\item[(ii)] \emph{Freedom of choice}: the I-bits and the D-bit can only be correlated with events in its causal future; also the they are taken to be completely random, i.e., $a,b$, and $b'$ take values $0$ or $1$ with probability $1/2$;
\item[(iii)] \emph{Closed laboratories}: each party's guess can be correlated with the bit generated by other party only if the latter is generated in the causal past of the system entering to the guessing party's laboratory.
\end{itemize}

Here we assume that Alice and Bob generate their bit $a$ and $b$, respectively, according to a biased coin with probability distribution $p(head)=\alpha$ and $p(tail)=1-\alpha$. Outcome of the coin `head' (`tail') corresponds to the bit value $`0'(`1')$. Bob uses another biased coin with probability distribution $p(head)=\beta$ and $p(tail)=1-\beta$ to generate the bit $b'$. Without loss of generality we assume that $\frac{1}{2}\le\alpha,\beta<1$. This can be understood as a relaxation of the freedom of choice assumption. The probability of success of this biased game is
\begin{eqnarray}\label{bia_succ}
p_{succ}(bias)&:=&\beta p(x=b|b'=0)\nonumber\\
&&~~~~+(1-\beta)p(y=a|b'=1).
\end{eqnarray}      
Now for the bias $(\alpha,\beta)$ if all events obey causal order, the two fellows cannot exceed the bound: 
\begin{equation}\label{bias_causal}
p^{csl}_{succ}(\alpha,\beta)\le\beta+\alpha(1-\beta).
\end{equation}
If all events satisfy deterministic causal order then only one way communication is possible (either from Alice to Bob or from Bob to Alice). Consider that Alice's operation follows Bob's one. Then Bob can send his bit $b$ to Alice and they will perform their task perfectly whenever $b'= 0$. However, when Bob is asked to guess Alice's bit, he cannot do it better than guessing $0$ (since $0$ occurs more frequently than $1$, i.e. $\alpha\ge\frac{1}{2}$). This results in an overall success probability of $\beta+\alpha(1-\beta)$. Since $\beta\ge\frac{1}{2}$, the other way communication cannot do better. Even if we assume a probabilistic causal order, i.e., there is a deterministic causal order but unknown due to \emph{subjective} ignorance, still one cannot overcome this bound. The formal proof of the BCI (\ref{bias_causal}) completely resembles the original proof of Oreshkov \emph{et al.}\cite{Oreshkov'2012}, but for completeness we present it in Appendix-\ref{appen-A}.

The inequality (\ref{bias_causal}) contains inequality (\ref{causal}) as a special case, i.e., $p^{csl}_{succ}(\frac{1}{2},\frac{1}{2})\le\frac{3}{4}$. However, in the following sections we will study two interesting cases:
\begin{itemize}
\item Case-I: The I-bits are unbiased, i.e, $\alpha=\frac{1}{2}$, whereas the D-bit has bias $\beta$. In this case the causal inequality (\ref{bias_causal}) becomes,
\begin{equation}\label{biase-b'}
p^{csl}_{succ}(\frac{1}{2},\beta)\le\frac{1+\beta}{2}.
\end{equation}
\item Case-II: Here the D-bit is unbiased, i.e., $\beta=\frac{1}{2}$ but the I-bits have bias $\alpha$. In this case the causal inequality (\ref{bias_causal}) reads,
\begin{equation}\label{biase-a}
p^{csl}_{succ}(\alpha,\frac{1}{2})\le\frac{1+\alpha}{2}.
\end{equation}

\end{itemize}   

\section{Process matrix formalism}\label{sec3}
The most general scenario compatible with the assumption that the operations performed in each local laboratory are described by quantum theory is
conveniently represented in the process matrix formalism introduced by Oreshkov \emph{et al.}\cite{Oreshkov'2012} At this point it is important to note that Chiribella \emph{et al.} have independently introduced the comb formalism to describe causally ordered quantum networks \cite{Chiribella'2009} which has been further generalized beyond causally ordered networks \cite{Chiribella'2013}.

\subsection{Process matrix}
Alice's local quantum laboratory can be specified by an input Hilbert space $\mathcal{H}^{A_I}$ and an output Hilbert space $\mathcal{H}^{A_O}$ with dimensions $d_{A_I}$ and $d_{A_O}$, respectively. The most general local operation is described by a completely positive (CP), trace non-increasing map $\mathcal{M}^A_i:\mathcal{L}(\mathcal{H}^{A_I})\mapsto\mathcal{L}(\mathcal{H}^{A_O})$ \cite{Nielsen'2000}, where $\mathcal{L}(\mathcal{H}^{X})$ denotes the space of linear hermitian operators over the Hilbert space $\mathcal{H}^{X}$. An instrument \cite{Davies'1970} is defined as the collection $\{\mathcal{M}^A_i\}_{i-1}^m$ of CP maps satisfying the condition that $\mathcal{M}^A=\sum_{i=1}^{m}\mathcal{M}^A_i$ a is CP and trace-preserving (CPTP). Choi-Jamio\l{}kowski (CJ) isomorphism provides an advantageous representation of CP maps by positive semi-definite matrices \cite{Jamio'1972,Choi'1975}. The CJ matrix corresponding to a linear map $\mathcal{M}^A_i:\mathcal{L}(\mathcal{H}^{A_I})\mapsto\mathcal{L}(\mathcal{H}^{A_O})$ is given by 
$$M^{A_IA_O}_i:=[\mathcal{I}\otimes\mathcal{M}^A_i(|\mathbb{1}\rangle\rangle\langle\langle\mathbb{1}|)]^T\in\mathcal{L}(\mathcal{H}^{A_I}\otimes\mathcal{H}^{A_O}),$$
where $\mathcal{I}$ is identity map, $|\mathbb{1}\rangle\rangle\equiv|\mathbb{1}\rangle\rangle^{A_IA_I}:=\sum_{k=1}^{d_{A_I}}|k\rangle^{A_I}\otimes|k\rangle^{A_I}\in\mathcal{H}^{A_I}\otimes\mathcal{H}^{A_I}$ is unnormalized maximally entangled state and $T$ denotes transposition in the bases $\{|k\rangle^{A_I}\}_{k=1}^{d_{A_I}}$. A map is completely positive \emph{iff} its CJ matrix is positive semidefinite and the trace-preserving condition is equivalent to $\mbox{Tr}_{A_O} M^{A_IA_O} = \mathbb{1}^{A_I}$, where $\mbox{Tr}_{X}$ denotes the partial trace over subsystem $X$, and $\mathbb{1}^{A_I}\in\mathcal{L}(\mathcal{H}^{A_I})$ is the identity matrix.

In the case of more than one party, the probability of obtaining local outcomes corresponding to a set of CP maps $\mathcal{M}^A_i,\mathcal{M}^B_j,..$, is a multi-linear function $P(\mathcal{M}^A_i,\mathcal{M}^B_j,..)$. Using this CJ representation, for two party scenario the bilinear function reads \cite{Oreshkov'2012},
\begin{equation}\label{gen_born}
P(\mathcal{M}^A_i,\mathcal{M}^B_j)=\mbox{Tr}\left[\mathcal{W}^{A_IA_OB_IB_O}\left( M^{A_IA_O}_i\otimes M^{B_IB_O}_j\right) \right], 
\end{equation} 
where $\mathcal{W}^{A_IA_OB_IB_O}\in\mathcal{L}(\mathcal{H}^{A_I}\otimes\mathcal{H}^{A_O}\otimes\mathcal{H}^{B_I}\otimes\mathcal{H}^{B_O})$. The requirement that $P(\mathcal{M}^A_i,\mathcal{M}^B_j)$ should be valid probability(for all possible choice of quantum operations including operations involving possibly entangled ancillary systems) put further constraints on $\mathcal{W}^{A_IA_OB_IB_O}$, which are 
\begin{eqnarray}\label{pross1}
\mathcal{W}^{A_IA_OB_IB_O}\ge 0,~~~~~~~~~~~~~~~~~~~~~~~~~~~~~~~~~~\\\label{pross2}
\mbox{Tr}\left[\mathcal{W}^{A_IA_OB_IB_O}\left( M^{A_IA_O}\otimes M^{B_IB_O}\right) \right]=1,~~~~~~~~~~~~~~~~~\\
\forall M^{A_IA_O}, M^{B_IB_O}\ge 0; \mbox{Tr}_{A_O}M^{A_IA_O} =\mathbb{1}^{A_I};~~~~~~~\nonumber\\ \mbox{Tr}_{B_O}M^{B_IB_O} =\mathbb{1}^{B_I}. ~~~~~~~~~~~  \nonumber
\end{eqnarray}   
A matrix $\mathcal{W}^{A_IA_OB_IB_O}$ that satisfies these conditions is called a process matrix. This generalizes the notion of quantum state and Eq.(\ref{gen_born}) can be thought of as generalization of Born rule in this formalism.

\subsection{Causally separable processes and causal correlations} 
As pointed out by Oreshkov \emph{et al.} \cite{Oreshkov'2012}, a causal structure is a set of event locations equipped with a partial order $\preceq$ that defines the possible directions of signalling. Whenever $A\preceq B$, reads as \emph{`$A$ is in the causal past of $B$'} or \emph{`$B$ is in the causal future of $A$'}, signalling from $A$ to $B$ is possible, but not from $B$ to $A$. The most general situation in quantum theory where signalling from Alice to Bob is not possible can be described by quantum channel with memory. In such a case Bob operates on one part of an entangled state and his output plus the other part are transferred to Alice through a channel. Process matrices of this kind will be denoted by $\mathcal{W}^{A\npreceq B}$. It may happen that $B\npreceq A$ occurs with probability $0\le q\le 1$ and rest of the time $A\npreceq B$ occurs. Such situation can be represented by a process matrix of the form
\begin{equation}\label{causal-process}
\mathcal{W}^{A_IA_OB_IB_O}=q\mathcal{W}^{B\npreceq A}+(1-q)\mathcal{W}^{A\npreceq B}.
\end{equation}
This can be thought as events are ordered with respect to some definite causal relation but there is a subjective ignorance about the particular order they follow. Process matrices of this form will not violate any of the causal inequalities (\ref{causal}), (\ref{bias_causal}), (\ref{biase-b'}), or (\ref{biase-a}).

Causal inseparability can be device-independently confirmed using \emph{causal inequalities} (this is analogous to the fact that entanglement of a quantum state can be certified \emph{device-independently} if the probability distribution originating from a set of measurements violates a Bell inequality), where the ``non-causal'' correlations between Alice and Bob alone suffice to show that the process they use is not causally separable, without additional trust in their local operations.

The condition for a probability distribution to be ``causal'', i.e., \emph{not} to violate any causal inequality, is simply that it can be decomposed into a convex combination of a probability distribution which is no-signaling from Bob to Alice ($p_{A \prec B}$) and a probability distribution which is no-signaling from Alice to Bob ($p_{B \prec A}$)\cite{Branciard'2015}:
\begin{equation}\label{eq:causal-prob}
p_{\text{causal}} = q p_{A \prec B} + (1-q) p_{B \prec A}.
\end{equation}
Thus the correlations generated by a causally ordered process(\ref{causal-process}) cannot violate any of the causal inequalities (\ref{causal}), (\ref{bias_causal}), (\ref{biase-b'}), or (\ref{biase-a}).

\section{Violation of BCI by causal inseparable process}\label{sec4}
There exists valid process matrix that violates the causal inequality (\ref{causal}). This essentially means that the corresponding process matrix is \emph{causally inseparable}, i.e., can not be expressed in a causal separable form. In this section we will discuss the possibility of violating the BCIs (\ref{biase-b'}) and (\ref{biase-a}) with valid inseparable processes.
\subsection{Case (I): Biased D-bit}
First consider the causal inequality (\ref{biase-b'}). Any causally separable process satisfies the bound
\begin{equation*}
p^{csl}_{succ}(\frac{1}{2},\beta)\le\frac{1+\beta}{2},
\end{equation*} 
where $\frac{1}{2}\le\beta<1$. Consider now, the following process matrix
\begin{equation}\label{casul-insep(b')}
\mathcal{W}^{A_IA_OB_IB_O}_{\beta}=\frac{1}{4}\left[\mathbb{1} + f_1(\beta)\sigma_z^{A_O}\sigma_z^{B_I} +f_2(\beta) \sigma_z^{A_I}\sigma_x^{B_I}\sigma_z^{B_O}\right].
\end{equation}
where $f_1(\beta)=\frac{1-\beta}{\sqrt{1-2\beta+2\beta^2}}$, $f_2(\beta)=\frac{\beta}{\sqrt{1-2\beta+2\beta^2}}$. It is straightforward to verify that for the said range of $\beta$, $\mathcal{W}^{A_IA_OB_IB_O}_{\beta}$ are valid process matrices, i.e., satify the conditions (\ref{pross1}) and (\ref{pross2}). Alice and Bob apply the following CP maps (\ref{alice-cp}) and (\ref{bob-cp}), respectively,
\begin{equation}\label{alice-cp}
\xi^{A_IA_O}(x,a)=\frac{1}{4} \left[\mathbb{1}+(-1)^x\sigma_z \right]^{A_I}\otimes\left[\mathbb{1}+(-1)^a\sigma_z \right]^{A_O},
\end{equation}
and the CJ matrix of the CP map performed by Bob is given by
\begin{equation}
\label{bob-cp}
\eta^{B_IB_O}(y,b,b')=  b'\eta_1^{B_IB_O}(y,b)+(b'\oplus 1)\eta_2^{B_IB_O}(y,b), 
\end{equation}
where,
$$\eta_1^{B_IB_O}(y,b) =  \frac{1}{2} \left[\mathbb{1} + (-1)^y\sigma_z  \right]^{B_I}\otimes\rho^{B_O},$$  
$$\eta_2^{B_IB_O}(y,b) = \frac{1}{4}  \left[\mathbb{1} + (-1)^y\sigma_x  \right]^{B_I}\otimes\left[\mathbb{1} + (-1)^{b\oplus y}\sigma_z\right]^{B_O}.$$

The joint conditional probability is determined as, $$p(xy|abb')=\mbox{Tr}\left[\mathcal{W}^{A_IA_OB_IB_O}_{\beta}
\left( \xi^{A_IA_O}(x,a)\otimes\eta^{B_IB_O}(y,b,b')\right) \right].$$ 
And we have [see Appendix-\ref{appen-B} for details],
\begin{eqnarray*}
p(x|ab,b'=0)&=&\sum_yp(xy|ab,b'=0)\nonumber\\
&=&\frac{1}{2}\left(1+(-1)^{x\oplus b}f_2(\beta) \right), 
\end{eqnarray*}   
which further implies, 
\begin{equation}\label{a}
p(x=b|b'=0)=\frac{1}{2}\left(1+f_2(\beta) \right).
\end{equation}
Similarly, we have,
\begin{eqnarray*}
p(y|ab,b'=1)&=&\sum_xp(xy|ab,b'=1)\nonumber\\
&=&\frac{1}{2}\left(1+(-1)^{y\oplus a}f_1(\beta) \right),
\end{eqnarray*}
implying, 
\begin{equation}\label{b}
p(y=a|b'=1)=\frac{1}{2}\left(1+f_1(\beta)\right).
\end{equation}
Replacing Eqs. (\ref{a}) and (\ref{b}) in the right hand side of Eq.(\ref{bia_succ}) we get,
\begin{eqnarray}\label{max1}
p_{succ}(\mathcal{W}_{\beta})&=&\beta\frac{1}{2}(1+f_2(\beta))+(1-\beta)\frac{1}{2}(1+f_1(\beta))\nonumber\\
&=&\frac{1}{2}(1+\sqrt{1-2\beta+2\beta^2}).
\end{eqnarray}
\begin{figure}[t]
\centering 
\includegraphics[width=8cm,height=5cm]{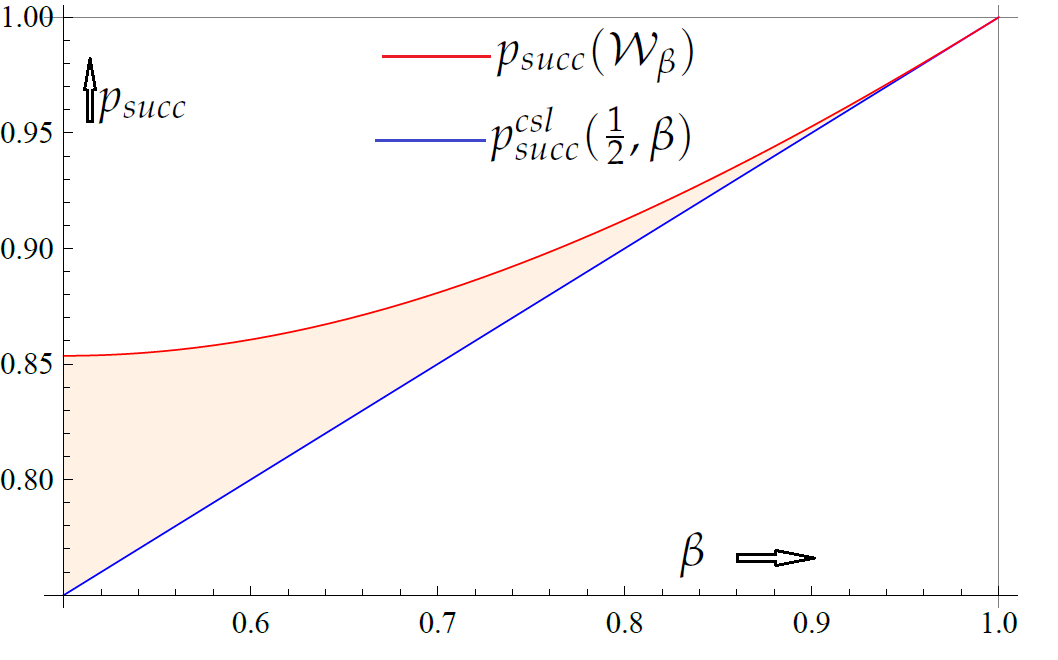} 
\caption{(Color on-line) Throughout the range of $\beta$, $p_{succ}(\mathcal{W}_{\beta})>p_{succ}^{csl}(\frac{1}{2},\beta)$. However the gap decreases with increasing $\beta$.}
\label{fig1} 
\end{figure}
Clearly, $p_{succ}(\mathcal{W}_{\beta})>p_{succ}^{csl}(\frac{1}{2},\beta)$ [see Fig.\ref{fig1}]. Moreover in the following section we show that if Alice and Bob apply local operations involving traceless binary observables then using most general valid bipartite process matrix the maximal value of the causal inequality (\ref{biase-b'}) is upper bounded by $\frac{1}{2}(1+\sqrt{1-2\beta+2\beta^2})$.  

Also one can construct \emph{causal witness} that detect causal inseparability of the process matrix of Eq.(\ref{casul-insep(b')}). The concept of causal witness \cite{Araujo'2015} is analogous to the ides of entanglement witness \cite{Horodecki'2009,Guhne'2009}. A hermitian operator $S_{\mathcal{W}_{ns}}$ can witness the causally inseparable process matrix $\mathcal{W}_{ns}$ if, 
$$\mbox{Tr}[S_{\mathcal{W}_{ns}}\mathcal{W}^{sep}]\ge0,~\forall~\mathcal{W}^{sep}; \& ~\mbox{Tr}[S_{\mathcal{W}_{ns}}\mathcal{W}_{ns}]<0;$$ 
where $\mathcal{W}^{sep}$ are the causally separable process matrix. The hermitian operator $S=\frac{1}{4}\left[\mathbb{1} -\sigma_z^{A_O}\sigma_z^{B_I}-\sigma_z^{A_I}\sigma_x^{B_I}\sigma_z^{B_O}\right]$ is a valid witness for the inseparable process matrix (\ref{casul-insep(b')}) as,
 $$\mbox{Tr}[S\mathcal{W}_{\beta}]=1-\frac{1}{\sqrt{1-2\beta+2\beta^2}}<0, \forall\beta,$$
and $\mbox{Tr}[S\mathcal{W}^{sep}]\ge0$ for all causally separable processes \cite{Araujo'2015}. 
\subsection{Case (II): Biased I-bit}
As already discussed, whenever the D-bit is random but input bits are biased, the causal inequality (\ref{bias_causal}) turns out to be
\begin{equation*}
p^{csl}_{succ}(\alpha,\frac{1}{2})\le\frac{1+\alpha}{2}.
\end{equation*}
All separable process matrices satisfy this inequality. The causally inseparable process matrix,
\begin{equation}\label{casul-insep}
\mathcal{W}^{A_IA_OB_IB_O}=\frac{1}{4}\left[\mathbb{1} + \frac{1}{\sqrt{2}}\left(\sigma_z^{A_O}\sigma_z^{B_I} + \sigma_z^{A_I}\sigma_x^{B_I}\sigma_z^{B_O} \right) \right],
\end{equation}
violates this inequality as far as the bias parameter $\alpha<\frac{1}{\sqrt{2}}$. So naturally one, likewise previous scenario [Case(I)], can try to find out a parametric class of valid process matrices that violate the inequality throughout the parameter range $\alpha$. However, in the following we show that if Alice and Bob apply \emph{measurement-repreparation} type of local operations involving traceless binary observables then no bipartite qubit process matrix can violate this inequality whenever $\alpha>\frac{1}{\sqrt{2}}$. 
 
\section{Maximal violation of BCI}\label{sec5}
The analysis of this section is similar to the Ref.\cite{Brukner'2015} and for convenience we use analogous notations. A Hilbert-Schmidt basis of $\mathcal{L}(\mathcal{H}^X)$ is given by a set of matrices $\{\sigma^X_{\mu}\}_{\mu=0}^{d_X^2-1}$ with $\sigma^X_0=\mathbb{1}_X$, $\mbox{Tr}\sigma^X_{\mu}\sigma^X_{\nu}=d_X\delta_{\mu\nu}$, and $\mbox{Tr}\sigma^X_{j}=0$ for $j=1,...,d_X^2-1$, where $d_X$ is dimension of the Hilbert space $\mathcal{H}^X$. A traceless dichotomic observable $O_{\vec{m}}$ can be expressed as $O_{\vec{m}}=\sum_{i>0}m_i\sigma^X_i\equiv(\vec{m}|\vec{\sigma})$, where $\vec{m}$ completely specifies the observable and $O_{\vec{m}}^2=\mathbb{1}$ (strictly speaking, the fact that the observable is dichotomic does not imply that the square of the observable is identity(the eigenvalues could be 0 and 1, instead of $1$ and $-1$). However, it is always possible to re-define a dichotomic observable so that it is true that it squares to identity as mentioned in \cite{Brukner'2015}) implies one has $|\vec{m}|=1$. The projectors onto two eigen spaces are given as $P_{\vec{m}}^x=\frac{1}{2}[\mathbb{1}+(-1)^xO_{\vec{m}}]$ with outcomes $x = 0$ or $1$. A traceless dichotomic correlation observable is given by  $O_{\hat{T}}=\sum_{ij>0}T_{ij}\sigma^X_i\otimes\sigma^X_j$, where $O_{\hat{T}}^2=\mathbb{1}$ and $|\hat{T}|= 1$.

As noted in \cite{Brukner'2015}, the CJ matrix representation of Alice's possible operations involving traceless binary observables read $\xi^{A_IA_O}(x,a)=\frac{1}{2d_{A_O}}[\mathbb{1}+(-1)^x(\vec{m}|\vec{\sigma}^{A_I})+(-1)^F(\vec{n}|\vec{\sigma}^{A_O})+(-1)^{x\oplus F}(\hat{T}|\vec{\sigma}^{A_I}\otimes\vec{\sigma}^{A_O})]$, where encoding function $F(x, a)\in\{0,1\}$. Similarly CJ matrix of Bob's CP map is given by $\eta^{B_IB_O}(y,b,b')=b'\eta^{B_IB_O}_1(y,b)_1+(1\oplus b')\eta^{B_IB_O}_2(y,b)$, where $\eta^{B_IB_O}_1(y,b)=\frac{1}{2}[\mathbb{1}+(-1)^y(\vec{r}|\vec{\sigma})]^{B_I}\otimes\rho^{B_O}$, with $\rho^{B_O}$ an arbitrary state of $B_O$ and  $\eta^{B_IB_O}(y,b,b')=\frac{1}{2d_{B_O}}[\mathbb{1}+(-1)^y(\vec{t}|\vec{\sigma}^{B_I})+(-1)^G(\vec{o}|\vec{\sigma}^{B_O})+(-1)^{y\oplus G}(\hat{S}|\vec{\sigma}^{B_I}\otimes\vec{\sigma}^{B_O})]$, where encoding function $G(y,b)\in\{0,1\}$. The most general bipartite process is of the form
\begin{align*}
	\mathcal{W}^{A_IA_OB_IB_O} 	&= \frac{1}{d_{A_I}d_{B_I}}\left(\mathbb {1} + \sigma^{B\preceq A}+ \sigma^{A\preceq B}+\sigma^{A\npreceq\nsucceq B}\right) ,\\
	\sigma^{B\preceq A}	&:= \sum_{ij>0}c_{ij}\sigma^{A_I}_{i}\sigma^{B_O}_{j} + \sum_{ijk>0}d_{ijk}\sigma^{A_I}_{i}\sigma^{B_I}_{j}\sigma^{B_O}_{k},\\
	\sigma^{A\preceq B}	&:= \sum_{ij>0}e_{ij}\sigma^{A_O}_{i}\sigma^{B_I}_{j} + \sum_{ijk>0}f_{ijk}\sigma^{A_I}_{i}\sigma^{A_O}_{j}\sigma^{B_I}_{k},\\
	\sigma^{A\npreceq\nsucceq B} 	&:= \sum_{i>0}v_{i}\sigma^{A_I}_{i} + \sum_{i>0}x_{i}\sigma^{B_I}_{i} + \sum_{ij>0}g_{ij}\sigma^{A_I}_{i}\sigma^{B_I}_{j},\\
	\mbox{where }&	c_{ij}, d_{ijk}, e_{ij}, f_{ijk}, g_{ij}, v_{i}, x_{i} \in \mathbb {R},
\end{align*} 
with the condition $\mathcal{W}^{A_IA_OB_IB_O}\ge 0$. The joint conditional probabilities $P(xy|abb')$ are given by $P(xy|abb')=\mbox{Tr}\left[ \mathcal{W}^{A_IA_OB_IB_O}\left(\xi^{A_IA_O}(x,a)\otimes\eta^{B_IB_O}(y,b,b') \right) \right]$. However after calculation one has either
\begin{eqnarray}
P(x=b|a,b,b'=0)~~~~~~~~~~~~~~~\nonumber\\\label{a1}
=\frac{1}{2}\left[ 1+(-1)^b(\vec{v}|\vec{m})+(\hat{c}|\vec{m}\otimes\vec{o})\right],~\mbox{or}\\\label{a2}
P(x=b|a,b,b'=0)~~~~~~~~~~~~~~~\nonumber\\
=\frac{1}{2}\left[ 1+(-1)^b(\vec{v}|\vec{m})+(\hat{d}|\vec{m}\otimes\hat{S})\right].~~~~  
\end{eqnarray} 
Similarly one has either
\begin{eqnarray}
P(y=a|a,b,b'=1)~~~~~~~~~~~~~~~\nonumber\\\label{b1}
=\frac{1}{2}\left[ 1+(-1)^a(\vec{x}|\vec{r})+(\hat{e}|\vec{n}\otimes\vec{r})\right],~\mbox{or}\\\label{b2}
P(y=a|a,b,b'=1)~~~~~~~~~~~~~~~\nonumber\\
=\frac{1}{2}\left[ 1+(-1)^a(\vec{x}|\vec{r})+(\hat{f}|\hat{T}\otimes\vec{r})\right].~~~~  
\end{eqnarray}
Here the $\pm$ sign in front of the third terms on the right-hand side has been absorbed by a suitable choice of the vectors representing local operations. The components of $\hat{c},\hat{d},\hat{e},\hat{f},\vec{v},\vec{x}$ are respectively $c_{ij},d_{ijk},e_{ij},f_{ijk},v_{i},x_{i}$ as in the expression of the process matrix. 
\subsection{Biased D-bit inequality}
We perform our analysis with combination of the expressions Eq.(\ref{a2}) and Eq.(\ref{b1}). However the analysis is similar for other combinations of expressions. Since $\alpha=\frac{1}{2}$, we have, 
\begin{eqnarray*}
P(x=b|b'=0)&=&\frac{1}{2}\left[ 1+(\hat{d}|\vec{m}\otimes\hat{S})\right],~\mbox{and}\\
P(y=a|b'=1)&=&\frac{1}{2}\left[ 1+(\hat{e}|\vec{n}\otimes\vec{r})\right], 
\end{eqnarray*}  
which further imply,
\begin{equation}\label{bias-beta}
p_{succ}(\mathcal{W})=\frac{1}{2}+\frac{1}{2}[\beta(\hat{d}|\vec{m}\otimes\hat{S})+(1-\beta)(\hat{e}|\vec{n}\otimes\vec{r})].
\end{equation}
Therefore,
\begin{equation}
p^{max}_{succ}(\mathcal{W})=\frac{1}{2}+\frac{1}{2}\max_{\vec{m},\hat{S},\vec{n},\vec{r}}[\beta(\hat{d}|\vec{m}\otimes\hat{S})+(1-\beta)(\hat{e}|\vec{n}\otimes\vec{r})].
\end{equation}
Consider the vectors $\vec{a}\equiv(\mathbb{1}\otimes O_{\vec{n}}\otimes O_{\vec{r}}\otimes \mathbb{1})\sqrt{\rho}$, $\vec{b}\equiv(O_{\vec{m}}\otimes\mathbb{1}\otimes O_{\hat{S}} )\sqrt{\rho}$, and $\vec{i}\equiv(\mathbb{1}\otimes \mathbb{1}\otimes \mathbb{1}\otimes \mathbb{1})\sqrt{\rho}$, where $\rho=\frac{1}{d_{A_O}d_{B_O}}\mathcal{W}^{A_IA_OB_IB_O}$. The scalar product between two vectors $\vec{c}\equiv C$ and $\vec{d}\equiv D$ is defined by the inner product of the corresponding matrices, i.e., $(\vec{c}|\vec{d})\equiv\mbox{Tr}(C^{\dagger}D)$. With this inner product we have $(\vec{a}|\vec{a})=(\vec{b}|\vec{b})=(\vec{i}|\vec{i})=\mbox{Tr}\rho=1$, $(\vec{a}|\vec{i})=\mbox{Tr}[(\mathbb{1}\otimes O_{\vec{n}}\otimes O_{\vec{r}}\otimes \mathbb{1})\rho]=(\hat{e}|\vec{n}\otimes\vec{r})$, $(\vec{b}|\vec{i})=\mbox{Tr}[(O_{\vec{m}}\otimes\mathbb{1}\otimes \mathbb{1}\otimes O_{\vec{o}})\rho]=(\hat{d}|\vec{m}\otimes\hat{S})$, and $(\vec{a}|\vec{b})=0$. Thus we can write  
\begin{eqnarray*}
p^{max}_{succ}(\mathcal{W})&=&\frac{1}{2}+\frac{1}{2}\max_{\vec{a},\vec{b}}[\beta(\vec{b}|\vec{i})+(1-\beta)(\vec{a}|\vec{i})]\\
&=&\frac{1}{2}+\frac{1}{2}\max_{\phi}[\beta\cos\phi+(1-\beta)\sin\phi]\\
&\le&\frac{1}{2}(1+\sqrt{1-2\beta+2\beta^2}),
\end{eqnarray*}
which proves our claim.
\subsection{Biased I-bit inequality}
Since in this case $\alpha\ne\frac{1}{2}$, we have 
\begin{eqnarray*}
	P(x=b|b'=0)&=&\frac{1}{2}[1+(\hat{d}|\vec{m}\otimes\hat{S})]+\left( \alpha-\frac{1}{2}\right)(\vec{v}|\vec{m}),\\
	P(y=a|b'=1)&=&\frac{1}{2}[1+(\hat{e}|\vec{n}\otimes\vec{r})]+\left( \alpha-\frac{1}{2}\right)(\vec{x}|\vec{r}). 
\end{eqnarray*}
Since $\beta=\frac{1}{2}$, we have
\begin{eqnarray}
p^{max}_{succ}(\mathcal{W})&=&\max_{\vec{m},\hat{S},\vec{n},\vec{r}}\left\lbrace \frac{1}{4}[2+(\hat{d}|\vec{m}\otimes\hat{S})+(\hat{e}|\vec{n}\otimes\vec{r})]\right. \nonumber\\\label{bias-alpha}
&&\left. +\frac{1}{2}\left(\alpha-\frac{1}{2}\right)[(\vec{v}|\vec{m})+(\vec{x}|\vec{r})]\right\rbrace .
\end{eqnarray}

In qubit scenario consider \emph{measurement-repreparation} type of local operations as in \cite{Oreshkov'2012}. Alice measures the input system in some qubit basis and predicts $x$ according to measurement outcome. To encode $a$ she reprepares the system in the same basis and sends it as output system. Without loss of generality Alice can choose $z$ basis. Whenever $b'=1$, Bob measures the incoming system in a basis same as Alice's encoding basis (i.e $z$ basis) and guesses $y$ according to measurement result. In this case Bob's repreparation is irrelevant. For $b'=0$, Bob measures incoming system in some basis (say along $\vec{t}$ direction) and guesses $y$ accordingly. To encode $b$ he reprepares the qubit in a basis same as Alice's decoding basis (i.e $z$ basis). The CJ representation of their operations read         
\begin{eqnarray}\label{alice-cp1}
\xi^{A_IA_O}(x,a)&=&\frac{1}{4} \left[\mathbb{1}+(-1)^x\sigma_z \right]^{A_I}\otimes\left[\mathbb{1}+(-1)^a\sigma_z \right]^{A_O},~~\\\label{bob-cp1}
\eta^{B_IB_O}(y,b,b')&=& b'\eta_1^{B_IB_O}(y,b)+(b'\oplus 1)\eta_2^{B_IB_O}(y,b), 
\end{eqnarray}
where,
\begin{eqnarray*}\eta_1^{B_IB_O}(y,b) =  \frac{1}{2} \left[\mathbb{1} + (-1)^y\sigma_z  \right]^{B_I}\otimes\rho^{B_O},~~~~~~~~~~~~~~~\\  
\eta_2^{B_IB_O}(y,b) = \frac{1}{4}  \left[\mathbb{1} + (-1)^y\vec{t}.\vec{\sigma}\right]^{B_I}\otimes\left[\mathbb{1} + (-1)^{b\oplus y}\sigma_z\right]^{B_O}.
\end{eqnarray*}
The qubit process matrices contributing to the success probability for the above \emph{measurement-repreparation} type reads
\begin{eqnarray}\label{process1}
\mathcal{W}&=&\frac{1}{4}\left[\mathbb{1} + a_0\sigma_z^{A_O}\sigma_z^{B_I} + b_0\sigma_z^{A_I}\sigma_x^{B_I}\sigma_z^{B_O}+c_0\sigma_z^{A_I}\sigma_y^{B_I}\sigma_z^{B_O}\right. \nonumber\\
&&~~~~~~~~~~~~~~~\left. +d_0\sigma_z^{A_I}\sigma_z^{B_I}\sigma_z^{B_O}+ e_0\sigma_z^{A_I}+ f_0\sigma_z^{B_I}\right],
\end{eqnarray}
and the success probability turns out to be 
\begin{eqnarray}
P_{succ}(\mathcal{W})&=&\frac{1}{4}[2+a_0+t_1b_0+t_2c_0+t_3d_0\nonumber\\
&&~~~~~~~~~~~+(2\alpha-1)(e_0+f_0)].
\end{eqnarray}
Under the constraints $\mathcal{W}\ge 0$ and $\sum_{i=1}^3t_i^2=1$ we get $P^{max}_{succ}\le \frac{1}{4}(2+\sqrt{2})$.
 
One can still ask the question whether the success probability could be increased by exploiting general traceless binary observables on general bipartite process matrices. 

\section{Discussion}\label{sec6}  
The study of indefinite causal order is a new and interesting area of research. Apart from fundamental understanding of our physical world it possibly has many potential practical implications to be discovered in future. Though the study is different from the study of quantum nonlocality \cite{Bell'1964,Brunner'2014}, these two have some interesting connections. As shown in \cite{Brukner'2015}, the optimal bound (under a class of restricted instruments) of the causal inequality (\ref{causal}) is same as Cirel'son quantity \cite{Cirelson'1980}, which is the optimal upper bound of the Bell-Clauser-Horne-Shimony-Holt inequality \cite{Bell'1964,CHSH'1969}. On the other hand the authors in \cite{Baumeler'2013} have shown that perfect signaling correlations among three parties are possible which do not obey the restrictions imposed by global space-time and this result can be seen as an analog to a tripartite appearance of quantum non-locality as manifested in Greenberger-Horne-Zeilinger argument \cite{GHZ'1989} of quantum nonlocality. In the same direction, here we study the biased causal inequality. For biased decision bit we show that the BCI can always be violated by a valid process matrix and we also find the optimal violation under traceless observables. However, for biased input bits we show that this is not the case, i.e., there is a threshold value of bias beyond which no valid qubit process matrix can violate the BCI under \emph{measurement-repreparation} type of local operations and we conjecture this to be true for general scenario. In the case of nonlocality such biased scenario has been studied by Lawson \emph{et al.}\cite{Lawson'2010} \\

{\bf Acknowledgment:} We would like to gratefully acknowledge fruitful discussions with Guruprasad Kar, Arup Roy, and Amit Mukherjee. MB likes to thank Pinaki Parta and Md. Rajjak Gazi for numerous discussions. We also like to thank P. Perinotti for pointing out the Ref.\cite{Chiribella'2013}.  
 
\section{\normalsize{Appendix}}

\subsection{{Formal derivation of BCI (\ref{bias_causal})}}\label{appen-A}
In Ref.\cite{Oreshkov'2012}, causal relations are defined operationally. If an agent say Alice ($A$) can influence the outcomes of measurements performed by an agent say Bob ($B$), whereas $B$ is never able to influence $A$ unless $A=B$, then $A$ causally precedes $B$ by definition and denoted as $A\preceq B$ reads ``$A$ is in the \textit{causal past} of $B$", or equivalently, ``$B$ is in the \textit{causal future} of $A$". The relation $\preceq$ is a partial order means that it satisfies the following conditions: (i) \emph{reflexivity}: $A\preceq A$; (ii) \emph{transitivity}: if $A\preceq B$ and $B\preceq C$, then $A\preceq C$; and (iii) \emph{antisymmetry}: if $A\preceq B$ and $B\preceq A$, then $A=B$. The notation $A\npreceq B$ denotes $A$ is \textit{not} in the causal past of $B$. In a causal structure both $A\npreceq B$ and $B\npreceq A$ may hold which will be denoted by $A\npreceq\nsucceq B$.

The main events in the task are the systems entering Alice's and Bob's laboratories, which we will denote by $A_1$ and $B_1$, respectively, Alice generating the bits $a$ and $x$ (Alice's guess of Bob's bit $b$) and Bob generating the bits $b$, $b'$, and $y$ (Bob's guess of Alice's bit $b$). Alice generates the bit $a$ and produces her guess $x$ \textit{after} the system enters her laboratory, i.e., $A_1\preceq a,y$. Similarly, we have $B_1 \preceq b', b, y$. As already pointed out the assumptions behind the causal inequality are (i) \emph{definite causal structure}, (ii) \emph{free choice}, and (iii) \emph{Closed laboratories}. Whereas in the original derivation (Ref.\cite{Oreshkov'2012}) it is assumed that the bits $a$, $b$, and $b'$ are uniformly distributed, here we make these distributions biased. We assume that $p(a=0)=p(b=0)=\alpha$ and $p(b'=0)=\beta$ and without loss of generality $\frac{1}{2}\le\alpha,\beta\le 1$.  
  
In a causal structure we have $p (A_1\preceq B_1)+ p(B_1\preceq A_1)+p(A_1\npreceq\nsucceq B_1)= 1$, since these possibilities are mutually exclusive and exhaustive. From assumption \emph{FC} it follows that the bits $a$, $b$, and $b'$ are independent of which of these possibilities is realized. Using these facts, the success probability can be written
\begin{eqnarray}
p_{succ}=(1-\beta)p(y=a|b'=1)+\beta p(x=b|b'=0)\nonumber\\
= [(1-\beta)p(y=a|b'=1;A_1\preceq B_1)\nonumber \\
~~~+\beta p(x=b|b'=0;A_1\preceq B_1)]p(A_1\preceq B_1)\nonumber\\
+[(1-\beta)p(y=a|b'=1;B_1\preceq A_1)\nonumber \\
~~~+\beta p(x=b|b'=0;B_1\preceq A_1)]p(B_1\preceq A_1)\nonumber\\
+[(1-\beta)p(y=a|b'=1;A_1\npreceq\nsucceq B_1)\nonumber\\
+\beta p(x=b|b'=0;A_1\npreceq\nsucceq B_1)]p(A_1\npreceq\nsucceq B_1).~~~~~~\label{A1}
\end{eqnarray}
Let $A_1\preceq B_1$. Since $B_1\preceq b$, the transitivity of partial order implies $A_1\preceq b$. $x$ can only be correlated with $b$ if $b$ is in the causal past of $A_1$, thus $p(b| x; A_1\preceq B_1)= p(b| A_1\preceq B_1)$. Again $b$ is independent from the causal relations between $A_1$ and $B_1$ which implies  $p(b| A_1\preceq B_1)=p(b)$. Also $b$ and $b'$ are independent. We therefore have,
\begin{eqnarray}
p(x=b|b'=0;A_1\preceq B_1)=~~~~~~~~~~~~~~~~~~~~~~~~~~~~~~~~~\nonumber\\p(b=0; x=0|b'=0;A_1\preceq B_1)~~~~~~~~~~~~~~~~\nonumber\\
~~~~~~~~~~+p(b=1;x=1|b'=0;A_1\preceq B_1)~~~~~~~~~~\nonumber\\
= p(b=0| x=0; b'=0;A_1\preceq B_1)p(x=0| b'=0;A_1\preceq B_1)\nonumber\\+p(b=1|x=1; b'=0;A_1\preceq B_1) p( x=1| b'=0;A_1\preceq B_1)\nonumber\\
= \alpha p(x=0| b'=0;A_1\preceq B_1)~~~~~~~~~~\nonumber\\+ (1-\alpha)p ( x=1| b'=0;A_1\preceq B_1)\le \alpha.
~~~~~~~~~~~~~~~~~~~~~~~\label{A2}	
\end{eqnarray}	 
Similarly, when $B_1\preceq A_1$, we have $p(y=a|b'=1;B_1\preceq A_1)\le\alpha$; and if $A_1\npreceq\nsucceq B_1$, we obtain $p(y=a|b'=1;A_1\npreceq\nsucceq B_1)\le\alpha$ and $p(x=b|b'=0;A_1\npreceq\nsucceq B_1)\le\alpha$.
Substituting this in Eq.\eqref{A2}, we obtain
\begin{eqnarray}
p_{succ}\le [(1-\beta)p(y=a|b'=1;A_1\preceq B_1) 
+\beta \alpha]p(A_1\preceq B_1)\nonumber\\
+[(1-\beta)\alpha 
+\beta p(x=b|b'=0;B_1\preceq A_1)]p(B_1\preceq A_1)\nonumber\\
+\alpha p(A_1\npreceq\nsucceq B_1),~~~~~~~~~~~~~~~\nonumber\\
\le [(1-\beta)+\beta \alpha]p(A_1\preceq B_1) 
+[(1-\beta)\alpha 
+\beta ]p(B_1\preceq A_1) \nonumber\\
+\alpha p(A_1\npreceq\nsucceq B_1).~~~~~~~~~~~~~~~\label{A3}
\end{eqnarray}
For $\frac{1}{2}\le\alpha,\beta<1$, we have,
\begin{eqnarray*}
(1-\beta)+\beta\alpha&\le&(1-\beta)\alpha+\beta,\\
\alpha&\le&(1-\beta)\alpha+\beta.
\end{eqnarray*}
Substituting these in inequality (\ref{A3}) we obtain the required biased causal inequality (\ref{bias_causal}). 

\subsection{{Derivation of Eq. (\ref{max1})}}\label{appen-B}
We have $\sum_yp(xy|ab,b'=0)=\sum_y\mbox{Tr}\left[ \mathcal{W}^{A_IA_OB_IB_O}_{\beta}\left(\xi^{A_IA_O}(x,a)\otimes\eta_2^{B_IB_O}(y,b) \right) \right]$. Using linearity of Trace we can write, $$\sum_yp(xy|ab,b'=0)
=\mbox{Tr}_{A_IA_O}\left[\xi^{A_IA_O}(x,a)\mbox{Tr}_{B_IB_O}\left[ \mathcal{W}^{A_IA_OB_IB_O}_{\beta}\left( \sum_y\eta_2^{B_IB_O}(y,b) \right) \right]\right].$$
Since we have,
\begin{equation*}
\sum_y\eta_2^{B_IB_O}(y,b) = \frac{1}{2}  \left[\mathbb{1}^{B_IB_O} + (-1)^b\sigma_x^{B_I} \sigma_z^{B_O} \right], 
\end{equation*}
which implies,  
\begin{eqnarray*}
\mathcal{\bar{W}}^{A_IA_O}_{\beta}&=&\mbox{Tr}_{B_IB_O}\left[ \mathcal{W}^{A_IA_OB_IB_O}_{\beta}\left( \sum_y\eta_2^{B_IB_O}(y,b) \right) \right]\\
&=&\frac{1}{2}  \left[\mathbb{1}^{A_IA_O} + (-1)^bf_2(\beta) \sigma_z^{A_I} \right],
\end{eqnarray*}
and finally we get,
\begin{eqnarray*}
P(x|a,b,b'=0)&=&\mbox{Tr}_{A_IA_O}\left[\xi^{A_IA_O}(x,a)\mathcal{\bar{W}}^{A_IA_O}_{\beta}\right]\\
&=& \frac{1}{2}  \left[1+ (-1)^{b\oplus x}f_2(\beta)\right].
\end{eqnarray*}
Hence we have, $P(x=b|b'=0)=\frac{1}{2}  \left[1+ f_2(\beta)\right]$. Similarly we have,
$$\sum_xp(xy|ab,b'=1)
=\mbox{Tr}_{B_IB_O}\left[\eta_1^{A_IA_O}(y,b)\mbox{Tr}_{A_IA_O}\left[ \mathcal{W}^{A_IA_OB_IB_O}_{\beta}\left( \sum_x\xi^{A_IA_O}(x,a) \right) \right]\right].$$
Now using the fact that 
\begin{equation*}
\sum_x\xi^{A_IA_O}(x,a) = \frac{1}{2}  \left[\mathbb{1}^{A_IA_O} + (-1)^a \sigma_z^{A_O} \right], 
\end{equation*}
we get,  
\begin{eqnarray*}
\mathcal{\bar{W}}^{B_IB_O}_{\beta}&=&\mbox{Tr}_{A_IA_O}\left[ \mathcal{W}^{A_IA_OB_IB_O}_{\beta}\left( \sum_x\xi^{A_IA_O}(x,a) \right) \right]\\
&=&\frac{1}{2}  \left[\mathbb{1}^{B_IB_O} + (-1)^af_2(\beta) \sigma_z^{B_I} \right],
\end{eqnarray*}
and finally we have,
\begin{eqnarray*}
P(y|a,b,b'=1)&=&\mbox{Tr}_{B_IB_O}\left[\eta_1^{B_IB_O}(y,b)\mathcal{\bar{W}}^{B_IB_O}_{\beta}\right]\\
&=& \frac{1}{2}  \left[1+ (-1)^{a\oplus y}f_1(\beta)\right].
\end{eqnarray*}
Therefore, $P(y=a|b'=1)=\left[1+f_1(\beta)\right]$.

\end{document}